# A DFT based study of the low-energy electronic structures and properties of small gold clusters.

Prashant K. Jain

School of Chemistry and Biochemistry, Georgia Institute of Technology, Atlanta, GA 30332-0400, USA.

Gold clusters $Au_n$ of size $n$ = 2 to 12 atoms were studied by the density-functional theory with an ab-initio pseudopotential and a generalized gradient approximation. Geometry optimizations starting from a number of initial candidate geometries were performed for each cluster size, so as to determine a number of possible low-energy isomers for each size. Along with the lowest-energy structures, metastable structures were obtained for many cluster sizes. Interestingly, a metastable planar zigzag arrangement of Au atoms was obtained for every cluster size $n \geq 5$. The stable electronic structure, binding energy, relative stability and HOMO-LUMO gap for the lowest-energy isomer were calculated for each cluster size. Variation of the electronic properties with size is investigated in this paper and compared with experimental results and other calculations.

**Keywords**: Density-functional theory (DFT), gold nanoclusters, ab-initio.

## I. INTRODUCTION

Gold nanoclusters are a subject of much current research due to their important role as building blocks in nanoscale electronic,[1] optical and medical diagnostic devices.[2] Besides, the optical and electronic properties of nanosized clusters are also representative of how molecular scale behavior evolves into bulk solid-state phenomena.[3] In particular, small gold clusters have attracted interest as tips and contacts in molecular electronic circuits[4] and also as potential chemical catalysts.[5] However, before such practical applications can be developed, the chemical, thermodynamic, electronic and optical properties of the nanoclusters must be established. Apart from the most stable cluster isomers, other possible low-energy structures must also be investigated. However, direct and comprehensive experimental information of smaller gold clusters is difficult to acquire. Therefore, smaller gold clusters have been theoretically investigated using ab-initio computational methods over the past few years. Häkkinen and Landman[6] have investigated the low-energy structures of gold clusters and their anions for $n$ = 2 to 10 atoms using the density-functional theory (DFT) employing scalar-relativistic pseudopotentials for the $5d^{10}6s^1$ valence electrons of gold and a generalized gradient approximation (GGA). The same authors have also compared their DFT based electronic structure calculations of small gold cluster anions to experimental photoelectron spectroscopy (PES) data.[7] Based on similar ab-initio calculations, Häkkinen et al.[8] have investigated the interaction of gaseous $O_2$ with small gold clusters and anions. Yang et al.[9] have investigated the low-energy electronic structures and dielectric properties of small 2D gold clusters using the DFT method in the local density approximation (LDA). While several ab-initio calculations on gold nanoclusters in the past have been limited by symmetry constraints,[10] an unconstrained global minima search[11] is necessary for exhaustive search of minima on the cluster potential energy surface, so as to be able to predict all possible low-energy isomers. Considerable progress has been recently made towards this goal by GarzoÂn et al.,[12,13] who predicted amorphous structures for the most stable isomers of intermediate-size $Au_n$ ($n$=38, 55,75) nanoclusters based on an exhaustive minima search using genetic optimization methods that employed classical Gupta $n$-body potentials. Similarly, empirical potentials have been employed in the past to determine the lowest energy structures of medium-sized gold clusters.[14-17] However, recently Wang and Zhao[18] have used a combined scheme of DFT with empirical genetic algorithms to generate a number of possible structural isomers for gold clusters up to 20 atoms, which were further optimized at the DFT level to determine the lowest energy structures. In contrast to these earlier calculations, the work described in this paper uses a first-principles method based on DFT to predict a number of possible low-energy structural isomers for each cluster size $n$ = 2 to 12 atoms by carrying out unconstrained geometry optimizations starting from different initial candidate geometries, similar to the methodology used by

Häkkinen and Landman.[6] The computational method and the calculated results are discussed in the following paper.

## II. COMPUTATIONAL METHOD

Self-consistent-field (SCF) electronic structure calculations were carried out on small gold clusters $Au_n$ $n$ = 2 to 12 using the Q-Chem package.[19] The calculations were performed with the density-functional theory (DFT) in the generalized gradient approximation (GGA) using the Becke-3-Lee-Yang-Parr (B3LYP) functional [20,21] and the Stevens-Basch-Krauss-Jansien-Cundari (SBKJC) set[22] as the effective core potential (ECP), suitable for modeling some important relativistic effects in heavy atoms. Spin-orbit corrections were not included as per this calculation scheme. The calculation scheme employed the SBKJC basis set, which is a double numerical valence basis and includes a d-polarization function. The electronic structure was obtained by solving the Kohn-Sham equations self-consistently. Geometry optimizations were carried out with a convergence limit of $10^{-5}$ a.u. on the total electronic energy. The geometry optimizations were run starting from several initial candidate geometries, ranging from an open- structured arrangement of Au atoms to a close-packed structure. No symmetry constraints were imposed. Two different spin multiplicities were investigated for each cluster size: singlet/triplet for odd-numbered clusters and doublet/quartet for even-numbered clusters. The optimized electronic structure for each cluster was obtained from the Z-matrices in the Q-Chem program output while electronic properties were calculated from the SCF total electronic energy and the orbital energy values.

The accuracy of the chosen computational scheme was verified by a calculation of the ionization potential (IP) for the gold atom. The measured IP of the gold atom using the chosen scheme is 9.27 eV, which agrees well with the experimental value of 9.22 eV.[23] Substitution of SBKJC by the LANL2DZ ECP and basis yielded an IP value of 9.42 eV, whereas a local density approximation (LDA) resulted in an IP value of 8.78 eV. Similarly, the calculated value for the IP of $Au_5$ as per the chosen scheme is 8.25eV, which compares well with the experimental value of 8.00eV.[24]

## III. RESULTS AND DISCUSSIONS

### A. ELECTRONIC STRUCTURES

#### 1. Open-Packed Geometry: Metastable planar zigzag structures

The initial guess geometry turned out to be an important determining factor in the calculation of the optimized electronic structure, as depicted in Fig. 1. Initial guess geometries, in the proximity of an open structured arrangement of gold atoms, result in a metastable planar zigzag arrangement of gold atoms in a chain for each cluster size $n \geq 5$. However, the calculation does not predict any linear arrangement of gold atoms for $n \geq 5$. The average Au-Au bond distance for the planar zigzag structure is ca. 2.6 Å and the Au-Au-Au bond angles are ca. 126° for all cluster sizes. Häkkinen and Landman[7] predicted a similar metastable structure for $Au_4^-$ with an exactly similar Au-Au bond distance of 2.6 Å. This result is also similar to a DFT based study by SaÂnchez-Portal et al.,[25] which showed that gold monatomic wires exhibit a zigzag shape (with a bond distance of 2.32 Å and a bond angle of 131°) instead of a linear one. Only under extreme tension do the wires become linear, although this occurs just before the breaking point. Nevertheless, the ground state of the wire has a zigzag geometry with two Au atoms per unit cell. The reason for the chain taking a zigzag shape can be traced back to metallic character, which manifests itself in a tendency to increase the charge density by overlapping with neighbouring atoms so as to increase the binding energy.[26] The zigzag arrangement of atoms is thus stabilized due to the effect of symmetry-breaking, which lifts the two-fold degeneracy of a particular level so also increases the coordination of each atom from two to four nearest-neighbours. Zigzag arrangement of atoms has been experimentally observed in some nanowires of alkali metals[27] and is also a potential explanation of the unusually large bond distances measured in short Au nanowires.[28-30]

## 2. Close-Packed Geometry: Lowest-energy structures

Starting from close-packed geometries for each cluster size $n$, a number of low-energy metastable structures were obtained for each cluster size along with the most stable (lowest-energy) structure. The $Au_2$ dimer is calculated to have a bond length of 2.58 Å and an IR active vibration frequency of 176 cm$^{-1}$. These values compare well with DFT-LDA based calculations of Wang and Zhao (2.43 Å, 173 cm$^{-1}$)[18] and DFT-GGA calculations by Häkkinen and Landman (2.57 Å)[6] and are close to the experimental bond length value of 2.47 Å and frequency of 191 cm$^{-1}$ [31]. For the $Au_3$ trimer, a $D_{\alpha h}$ linear chain with a Au-Au bond distance of 2.64 Å is found to be more stable than a equilateral triangle with a side of 2.69 Å, similar to a result obtained by the DFT-LDA[18] which predicted a linear structure with a similar bond length of 2.67 Å as also earlier CASSCF studies. [32] Clusters containing 4 to 6 atoms have stable planar structures, as in case of other small metal clusters e.g. $Na_n$, $Ag_n$ and $Cu_n$ [33,34,35]. In case of $Au_4$, the most stable structure is a $D_{2h}$ rhombus, which is more stable than a square with a side of 2.69 Å and a "Y-shaped structure" by 0.69 eV and 1.54 eV respectively. $Au_5$ has an almost planar trapezoidal structure with $C_{2v}$ symmetry, which is more stable than a 3-D trigonal bipyramidal structure by about 1.5 eV. In accordance with the results of this calculation, earlier DFT-LDA[18] and DFT-GGA[6] calculations have both predicted the rhombus and the trapezoidal (or "W-shaped geometry") to be the most stable geometries for $Au_4$ and $Au_5$ respectively. The most stable geometry for $Au_6$ is predicted to be a hexagon with a side of 2.77 Å rather than a $D_{3h}$ planar triangle predicted by earlier calculations.[6,18] Photoelectron spectroscopy (PES) data show low electron affinity (EA) and large HOMO-LUMO gap for $Au_6$,[36] which is consistent with the properties of a hexagonal structure. $Au_7$ has a most stable 3-D pentagonal bipyramidal structure and a distorted metastable structure in agreement with Wang and Zhao.[18] Clusters of size 8 to 12 have complex irregular 3D structures. The calculated electronic structures show a transition from planar to non planar geometry at $n = 7$, similar to earlier DFT results,[18] though Häkkinen and Landman[6] have predicted non-planar structures from $n \geq 8$. Comparing with experiment, transition to non-planar cluster geometries has been determined to take place at $n = 7$ on the basis of ion mobility measurements for Au cluster cations.[37]

**B. ELECTRONIC PROPERTIES**

**1. HOMO-LUMO gap**

The energy gap between the highest occupied orbital (HOMO) and the lowest unoccupied orbital (LUMO) for the lowest energy isomers has been calculated for each cluster size. Transition from the atomic scale to bulk metallic behavior is accompanied by a closure of the HOMO-LUMO gap and development of collective electronic excitations.[38] A large HOMO-LUMO gap has also been considered as an important prerequisite for the chemical stability of gold clusters.[39] Figure 2 shows the size variation of this HOMO-LUMO gap ($E_g$ in eV). As seen from the plot, the HOMO-LUMO energy gap shows a decreasing trend with cluster size. This is consistent with the fact that, as molecular orbitals are formed from the overlap of more and more atomic orbitals, the energy levels progressively come closer.[38] In other terms, the valence electrons are delocalized over more number of atoms as the cluster size increases. The HOMO-LUMO gap values also show significant odd-even oscillations. The even-numbered clusters have a larger HOMO-LUMO gap than the neighbouring odd-sized ones. Odd-even alterations in the HOMO-LUMO gap values have been often predicted by earlier calculations for other small metal clusters[40,41] as also small $Au_n$ clusters.[6,9,18] Most importantly, HOMO-LUMO gaps evaluated from photoelectron spectroscopy (PES) data of gold cluster anions have also shown oscillations.[31,42] Odd-even oscillation behaviors can be understood by the effect of electron pairing in orbitals. Even-sized clusters have an even number of s valence electrons and a doubly occupied HOMO, while it is singly occupied for odd-sized clusters. The electron in a doubly occupied HOMO feels a stronger effective core potential due to the fact that the electron screening is weaker for electrons in the same orbital than for inner-shell electrons. It is crucial to note the unusually high values of the HOMO-LUMO gap at $n = 2$ and $n = 8$, unlike the HOMO-LUMO gaps obtained from earlier DFT calculations[6,18]. The values are however consistent with the well known fact that the HOMO-LUMO gap values are particularly large for $n = 2, 8, 20,…$, the magic numbers for electronic shell closing in small monovalent metal clusters.[43] It is possible, using other methods,[14-17] to calculate the HOMO-LUMO energy gap for larger gold clusters and thus study the closure of the gap and hence the onset of metallicity in gold nanostructures.

## 2. Binding energy and relative stability

The binding energy per atom was calculated for each cluster size $n$ as

$$E_B = [E_T - nE_{atom}]/n ,$$

where $E_T$ is the total electronic energy of a cluster and $E_{atom}$ is the total electronic energy for gold atom. The calculated binding energies are lower than those calculated by DFT-LDA.[18] However, it is known that GGA underestimates the binding energy by 0.3–0.4 eV/atom with respect to LDA based values.[6,44] As a comparison, this calculation yields a value of 0.95 eV/atom for $Au_2$ as against a value of 1.22 eV/atom[18] by DFT-LDA and an experimental value of 1.10 eV/atom.[45] Figure 3 gives the variation of the per atom binding energy $E_B$ with cluster size $n$. As seen in the plot, the per atom binding energy increases with cluster size. This trend is due to the increase of the average number of nearest-neighbours with increasing size, thus promoting greater average number of interactions per atom. This may also be an explanation for the higher binding energy of $Au_5$ with respect to the planar $Au_6$. Towards the higher values of $n$, the even-sized clusters show binding energies higher than their odd-sized neighbours. This is due to the effect of electron pairing as explained for the oscillations in the HOMO-LUMO gap. One can further calculate binding energies for clusters larger than 12 atoms using methods suitable for medium and large-sized clusters [14-17] and thus study how the value of the binding energy evolves into the bulk cohesive energy value of 3.90 eV/ atom.[46] Further, the second difference in energy for the clusters may be calculated as,

$$\Delta_2 E(n) = [E(n-1) + E(n+1) - 2E(n)],$$

where $E(n)$ is the total energy of the cluster of size $n$. In cluster physics, the quantity $\Delta_2 E(n)$ is commonly known to represent the relative stability of a cluster of size $n$ with respect to its neighbours.[41,43,47] Also, $\Delta_2 E$ can be directly compared to the experimental relative abundance: the peaks in $\Delta_2 E$ are known to coincide with the discontinuities in the mass spectra.[48] Figure 4 shows a plot of $\Delta_2 E$ as a function of the cluster size $n$. As seen, the plot shows odd-even alterations, similar to oscillations in the HOMO-LUMO gap, thus predicting a higher relative stability for even-sized clusters. However, a higher stability is predicted for n = 5 than n = 6, which may be due to the higher number of nearest neighbours per atom for $Au_5$ than for the planar $Au_6$. Also, magic-sized clusters $n$ = 2, 8 so also $n$ = 10 show a higher stability.

## IV. SUMMARY

Thus the low-energy geometries, binding energy, relative stability and HOMO-LUMO gap of gold clusters $Au_n$, $n$ varying from 2 to 12 atoms were investigated using a DFT based method. Table 1 summarizes the structure and properties of the lowest-energy isomer for each cluster size $n$. It was found that different initial guess geometries lead to a number of structural isomers for each cluster size. Most interestingly, for cluster sizes $n \geq 5$, open structured guess geometries predict an additional metastable planar zigzag arrangement of gold atoms. The calculated lowest energy structures show that the smallest gold clusters are planar and a transition to a 3D structure takes place at a size of 7 atoms. Odd-even alterations are evident in the HOMO-LUMO gaps and second energy differences. Even-numbered clusters show relatively higher stability. The calculated results are found to compare well with experimental data and earlier theoretical investigations of gold nanoclusters so also are consistent with theoretical concepts of magic numbers and shell models. However, due to the required computational effort, this first-principles method is appropriate only for small gold clusters.[49,50] Medium and large-sized clusters should be appropriately studied using methods employing empirical potentials,[14-17] to be able to accomplish a more thorough study of the size evolution of properties in gold nanostructures.

## ACKNOWLEDGEMENTS

I would like to thank Prof. C.David Sherill for his advice so also for providing the computational facilities at the CCMST, Georgia Tech for this work.

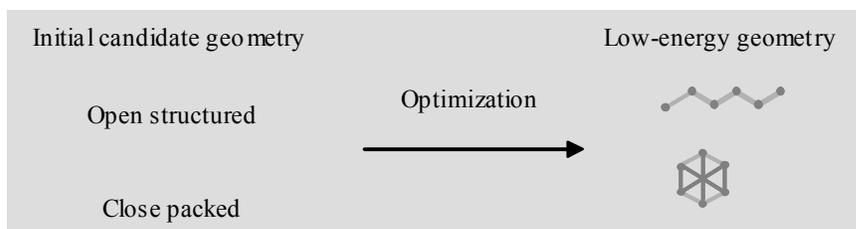

Fig.1. Initial candidate geometry determines the optimized low-energy geometry, for instance for *n* = 6

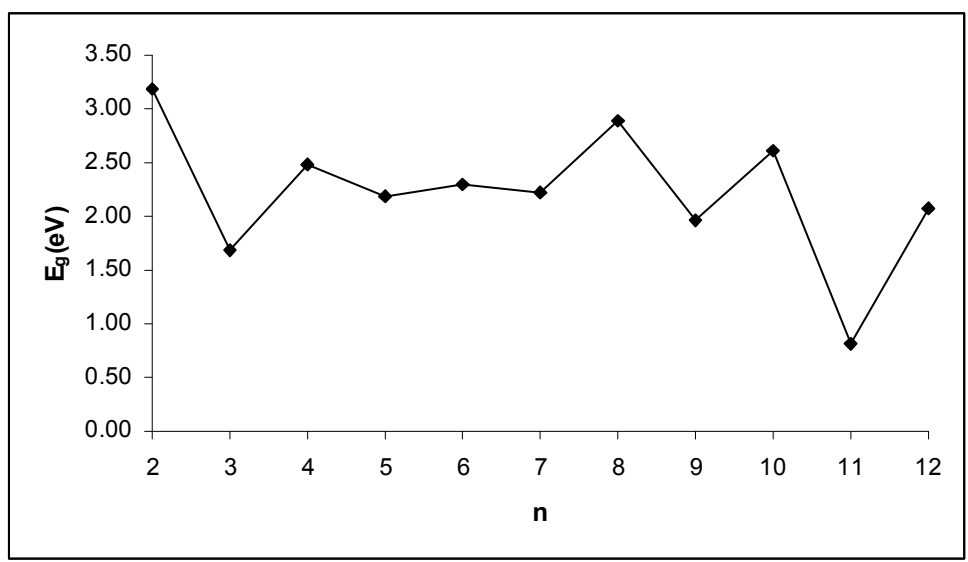

Fig.2. Plot of HOMO-LUMO gaps for different cluster sizes

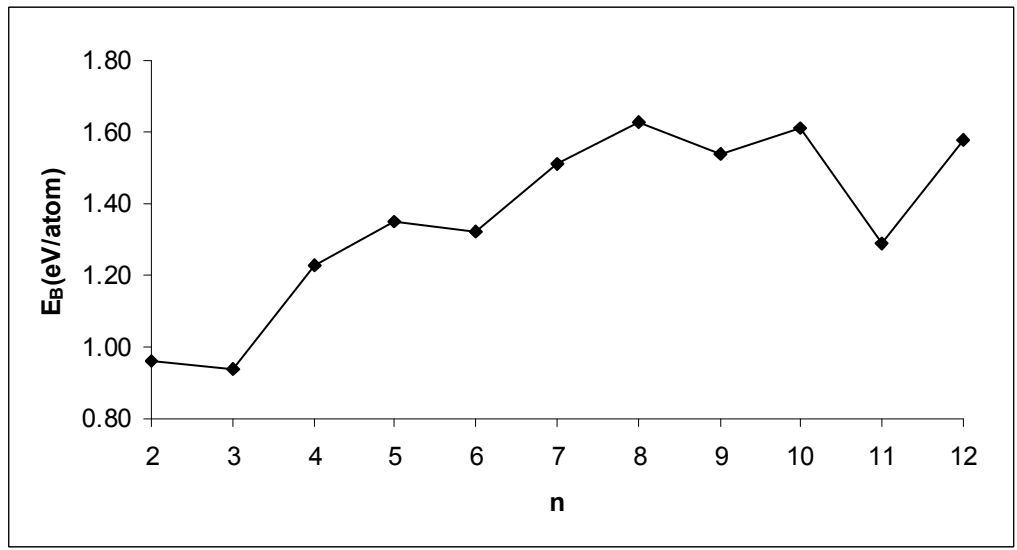

Fig.3. Binding energy/atom vs. cluster size *n*

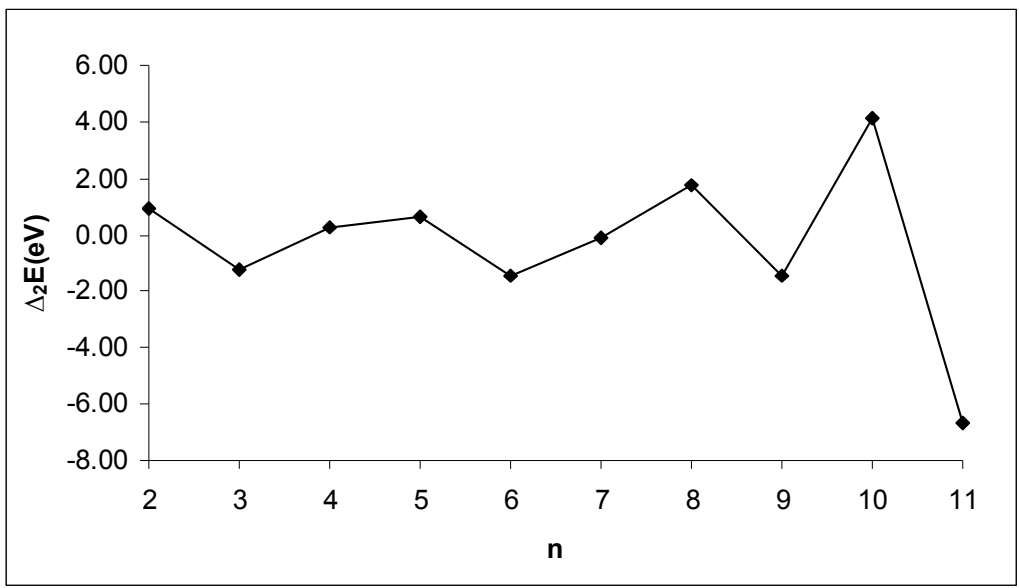

Fig.4. Second energy difference in total energy as a function of cluster size

| n | Stable geometry | | $E_g$ (eV) | $E_B$ (eV/atom) |
|---|---|---|---|---|
| 2 | dimer | 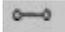 | 3.18 | 0.96 |
| 3 | linear chain | 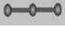 | 1.69 | 0.94 |
| 4 | rhombus | 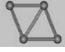 | 2.48 | 1.23 |
| 5 | trapezoidal | 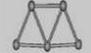 | 2.18 | 1.35 |
| 6 | hexagonal | 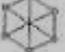 | 2.29 | 1.32 |
| 7 | pentagonal bipyramidal | 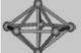 | 2.23 | 1.51 |
| 8 | irregular | | 2.88 | 1.63 |
| 9 | irregular | | 1.96 | 1.54 |
| 10 | irregular | | 2.61 | 1.61 |
| 11 | irregular | | 0.82 | 1.29 |
| 12 | irregular | | 2.07 | 1.58 |

Table 1. Stable geometry, HOMO-LUMO gap & binding energy/atom for $Au_n$, $n$ = 2-12